\documentclass[twocolumn]{elsart}

\usepackage{natbib}
\usepackage{graphicx}
\usepackage[a4paper]{hyperref}
\usepackage{amssymb}

\begin{document}

\begin{frontmatter}



\title{Parsec--scale radio jets in $\gamma$--ray loud sources}

 \author{Tiziana Venturi\corauthref{1}},
 \author{Daniele Dallacasa \corauthref{1}\corauthref{2}}
 \address[1]{Istituto di Radioastronomia, CNR, Via Gobetti 101, 40129 Bologna,
Italy}
 \address[2]{Dipartimento di Astronomia, Universit\`a di Bologna, 
Via Ranzani 1, 40127 Bologna}
 \ead{tventuri@ira.cnr.it}

\begin{abstract}
We will present a multiepoch study of the three blazars 0954+658 (BL--Lac), 
PKS\,1510--089 (HPQ) and 1749+096 (BL--Lac). The first two sources 
are known to be $\gamma$--ray loud. Our study is based on milliarcsecond 
resolution polarimetric observations carried out with the VLBA at 8.4 GHz. 
The observations took place between January 1999 and May 2001.
Superluminal motion is detected along the 
jet of PKS\,1510--089 and 1749+096, with $\beta_{app} \sim 10$
for all features. Magnetic field structure is revealed along
the jets of 0954+658 and PKS\,1510--089. The polarisation
properties of the parsec--scale jets remain stationary
in all sources, regardless of their total flux density variability
in the radio band and of the presence of superluminal features.
\end{abstract}

\begin{keyword}
Blazars: general \sep BL Lacertae objects \sep quasars: individual
(0954+658, PKS\,1510--089, 1749+096) \sep radio emission 
\end{keyword}

\end{frontmatter}


\section{Introduction}\label{sec1}

The blazar phenomenon is now believed to be the result
of an anisotropic relativistic jet propagating
at a small angle to the observer's line of sight.
Variability at all wavelengths and the
peculiar radio properties on the parsec scale, 
such as one--sidedness, jet bending, misalignement  
and superluminal features, can all be accounted for 
under these assumptions.\\
In a number of blazars the emission extends out to 
the MeV regime, i.e. to the $\gamma$--rays.
Many models have been proposed to explain the origin of 
this very high energy emission, such as for example
synchrotron self-Compton (i.e. Maraschi et al. 1992), 
inverse Compton scattering, with a variety of origins
for the scattered photons (i.e. Dermer et al. 1992; Sikora 
et al. 1994), synchrotron emission by ultra relativistic
electrons and positrons (Ghisellini et al. 1993). 
However, no consensus on the dominant process has been reached yet.\\
The connection between the $\gamma$--ray emission and the
parsec--scale radio properties is under study, and many 
questions are still unexplained. For example, it is
still unclear if (1) $\gamma$--ray loud blazars are
characterised by higher apparent speeds and/or higher 
Lorentz factors; (2) jet bending, wiggling and misalignement
are more prominent in $\gamma$--ray loud sources. Beyond this,
the correlation between radio variability and changes in
the magnetic field structure is still unclear for all classes
of radio loud AGNs.\\
In order to increase the statistics, and to shed light
on these important questions, detailed multifrequency studies 
for a high number of radio loud blazars is necessary.
In this paper we present a multiepoch study of three  
radio blazars, carried out at 8.4 GHz with the Very Long Baseline
Array (VLBA). Among our sources, two are also $\gamma$--ray loud.\\
We defined $h$=H$_{0}$/100 and $q_{0}=0.5$.

%
%
\section{Observations and data analysis}\label{sec2}\vspace{-5mm}
%

\begin{figure*}[t!]
  \centering
  \resizebox{11cm}{!}{\includegraphics{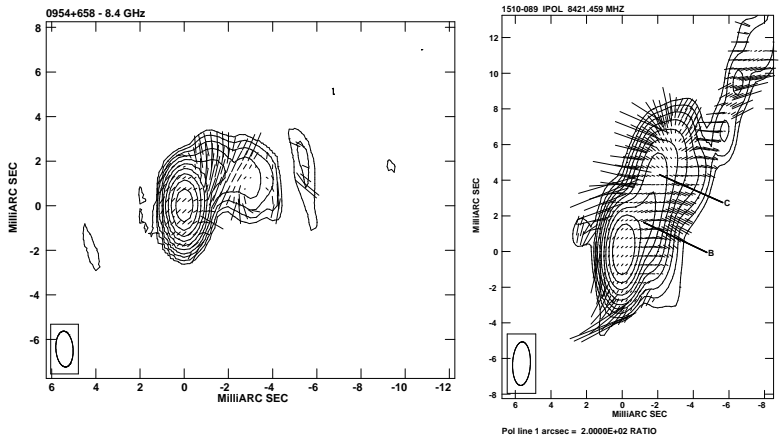}}
  \vspace{-5mm}
  \caption{\footnotesize{Left -- 
8.4 GHz image of 0954+658 (epoch December 2000).
Superposed vectors represent the fractional polarisation.
HPBW = 1.07$\times$0.65 mas, p.a. $14^{\circ}$.
Image peak 349 mJy/b. Contours are $-$0.4, 0.4,
0.8, 1.6, 3.2, 6.4, 12.8, 25.6, 50, 100, 200 mJy/b.
Right -- 8.4 GHz image of PKS\,1510--089 
(epoch December 2000).
The superposed vectors represent the fractional polarisation.
HPBW = 2.44$\times$1.00 mas, p.a. $-20^{\circ}$.
Image peak 726 mJy/beam. Contours are $-$0.6, 0.6,
1, 2, 4, 8, 16, 32, 62.5, 125, 250 mJy/beam.
}}
\label{Fig:0954}%
   \end{figure*}

\begin{figure}[t!]
  \centering
  \resizebox{6cm}{!}{\includegraphics{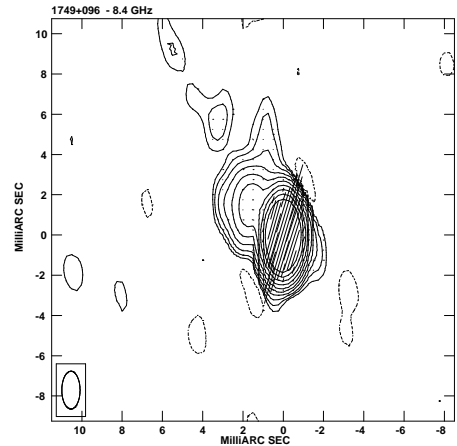}}
  \vspace{-5mm}
  \caption{\footnotesize{8.4 GHz image of 1749+096 (epoch December 2000).
The superposed vectors represent the magnetic field vectors.
HPBW = 1.88$\times$0.91 mas, p.a. $0^{\circ}$.
Image peak 3.40 Jy/b. Contours are $-$0.6, 0.6,
1.2, 2.4, 4.8, 9.6, 19.2, 38.4, 75, 150, 300 mJy/b.}} 
\label{Fig:1749}%
   \end{figure}

We performed multiepoch VLBA observations of the three blazars
0954+658 (LBL, z=0.368), PKS\,1510--089 (HPQ, z=0.361)
and 1749+096 (LBL, z=0.322). The first two sources have a
measured $\gamma$--ray flux (Hartman et al. 1999), while only
an upper limit exists for 1749+096 (Fichtel et al. 1994).
Snapshot observations were carried out
in January 1999, December 2000 and May 2001 in dual polarisation mode 
at 8.4 GHz and 22 GHz, for a total of 12 hours each run. The resulting 
total time on source was $\sim$ 2 hr/frequency. Here we will present 
and discuss the 8.4 GHz observations only.\\
The angular resolution of the VLBA at 8.4 GHz is $\sim$ 1 mas, and
the typical noise on the images is in the range 0.07 -- 0.3 mJy/beam,
depending on the peak flux density value.\\
Data calibration and imaging were carried out with the AIPS NRAO
package. The final datasets were then exported into the Difmap
package, where modelfitting of the visibilities was performed.
In Figs. \ref{Fig:0954} and \ref{Fig:1749} contour plots of
0954+658, PKS\,1510--089 and 1749+096 are given,
with superposed magnetic field vectors.

%
\section{The parsec--scale radio jets}\label{sec3}
%
All sources in our project showed considerable total flux density
variations during the period of our monitoring, 
up to $\sim$ 40\%.

\begin{figure}[t!]
  \centering
  \resizebox{6cm}{!}{\includegraphics{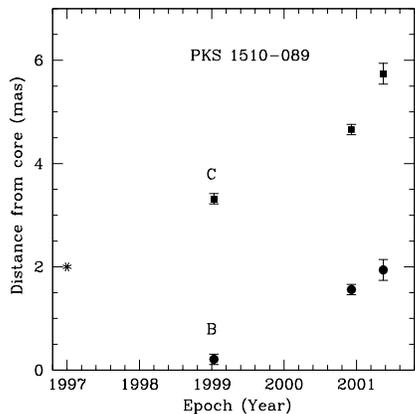}}
  \vspace{-5mm}
  \caption{\footnotesize{Position of the jet components in PKS\,1510--089.
The symbol $\star$ (epoch 1997) is taken from Homan et al. (2001)}}
\label{Fig:motion}%
   \end{figure}

We modelfitted the final visibilities at each epoch for
each source with gaussian components, in order to investigate
the presence of moving features.\\
PKS\,1510-089 shows major structural changes. The inner 6 mas are
well fitted with two components along the jet, labelled as B and C
in Fig. \ref{Fig:0954}, whose location with respect to the 
core (most likely associated with the image peak, based on
multifrequency analysis, Venturi et al. in prep.) increases with
time. The positions of B and C at the various epochs are
given in Fig. \ref{Fig:motion}. Here the position of component
C at the epoch 1997 is taken from Homan et al. (2001).\\
Both components are found to be superluminal, with a similar
value for the velocity, i.e.
$\beta_{app}$(B) = $10.0 \pm 0.3~h^{-1}$c, and 
$\beta_{app}$(C) = $10.4 \pm 1.2~h^{-1}$c. These values agree,
within the uncertainties, with the motion reported by Homan et
al. (2001), and imply a Lorentz factor $\gamma_{min} >$ 10.

The jet component in 1749+096 is also found superluminal,
with $\beta_{app} \sim 8.2~h^{-1}$c.

Our analysis suggests that the jet in 0954+658 is 
stationary. 
A comparison among our epochs and literature
images (Gabuzda \& Cawthorne 1996) leads either to no apparent motion, or 
to very high superluminal speed along the jet, with $\beta_{app} >$ 50c.
Kellermann et al. (2003) show that very high superluminal
features are found among the sources in the 2 cm VLBA survey,
however the case of 0954+658 would be very extreme.
We note that this source alternates core flux density flares
to long periods of quiescence (Venturi et al. 2001).

\medskip
A preliminary analysis of the polarised emission at 8.4 GHz 
shows that all sources have polarised flux in the 
core region and along the jet. The core polarised intensity
and polarisation angle change in all sources from
epoch to epoch. 
On the other hand, the polarised flux and the magnetic field orientation
along the jet remain stationary, regardless
of the flux density variability and of superluminal motion in the
knot.

In PKS\,1510--089 the magnetic field shows structure along the
jet. From Fig. {\ref{Fig:0954} (right panel) it is clear that
the fractional polarisation is higher in the external part of the jet,
while the jet spine is less polarised. This behaviour is found at
all epochs.

%
\section{Summary and conclusions}\label{sec4}
%
Our monitoring program on three blazars revealed different 
properties on the parsec--scale jets. In particular:

\noindent
-- superluminal motion is detected in PKS\,1510--089 (two
superluminal knots) and in 1749+096. In all cases an apparent
velocity $\beta_{app} \sim 10~h^{-1}$c is found;

\noindent
-- a stationary jet is present in 0954+658, despite the major 
variability in the radio band during the period of our monitoring
program;

\noindent
-- the polarisation properties along the parsec--scale jet
remain stationary from epoch to epoch in all sources, regardless
of their flux density variability and the presence of
superluminal features;

\noindent
-- magnetic field structure is found in PKS\,1510--089, where a
higher percentage of polarisation is detected along the outer edges
of the jet.

\bibliographystyle{aa}

\end{document}